\documentclass[preprint,showpacs,preprintnumbers,amsmath,amssymb,superscriptaddress]{revtex4}
\usepackage{amsmath}

\usepackage{graphicx}
\usepackage{dcolumn}
\usepackage{bm}
\usepackage{epsfig}

\begin{document}

\def\vv{{\hbox{\boldmath $v$}}}
\def\undersim#1{\setbox9\hbox{${#1}$}{#1}\kern-\wd9\lower
2.5pt \hbox{\lower\dp9\hbox to \wd9{\hss $_\sim$\hss}}}

\title{Two-pion interferometry for viscous hydrodynamic sources}

\author{M. J. Efaaf}
\affiliation{School of Physics and Optoelectronic Technology,
Dalian University of Technology, Dalian 116024, China}

\author{Zhong-Qian Su}
\affiliation{School of Physics and Optoelectronic Technology,
Dalian University of Technology, Dalian 116024, China}

\author{Wei-Ning Zhang}
\email[\ ]{wnzhang@dlut.edu.cn}
\affiliation{School of Physics and Optoelectronic Technology,
Dalian University of Technology, Dalian 116024, China}
\affiliation{Department of Physics, Harbin Institute of Technology, Harbin
150006, China}

\begin{abstract}
The space-time evolution of the (1+1)-dimensional viscous
hydrodynamics with an initial quark-gluon plasma (QGP) produced in
ultrarelativistic heavy ion collisions is studied numerically.  The
particle-emitting sources undergo a crossover transition from the
QGP to hadronic gas.  We take into account a usual shear viscosity
for the strongly coupled QGP as well as the bulk viscosity which
increases significantly in the crossover region.  The two-pion
Hanbury-Brown-Twiss (HBT) interferometry for the viscous hydrodynamic
sources is performed.  The HBT analyses indicate that the viscosity
effect on the two-pion HBT results is small if only the shear viscosity
is taken into consideration in the calculations.  The bulk viscosity
leads to a larger transverse freeze-out configuration of the
pion-emitting sources, and thus increases the transverse HBT radii.
The results of the longitudinal HBT radius for the source with
Bjorken longitudinal scaling are consistent with the experimental data.
\\[2ex]
\end{abstract}


\maketitle

\section{Introduction}

The experimental results of the Relativistic Heavy Ion Collider
(RHIC) at Brookhaven National Laboratory (BNL) indicate that the
matter produced in the central collisions of Au + Au at
$\sqrt{s_{NN}}=$130 and 200 GeV is a strongly coupled quark-gluon
plasma (sQGP), and behaves like a perfect liquid \cite{RHIC}.
Recently, the studies of the dissipative fluid dynamics become very
hot in high energy heavy ion collisions \cite{{MurRis04,Hei06,Bai06,
Cha06a,Cha06,Mur07,Rom07,Rom07L,Son08L,Son08,Son08a,Dus08,Luz08,
Luz09,Cha09,Mon09,Son09,Boz09,Li09}}, because people want to know
why the sQGP exhibits an almost perfect fluid property and how to
probe the viscosity effects on experimental observables.

The relativistic formulism of dissipative fluids was originally
derived by Eckart \cite{Eck40} and Landau and Lifshitz \cite{Lan59}.
Their theories contain only the first-order terms of the
dissipative quantities and therefore are referred to as the first-order
theories of dissipative fluids.  The theories of dissipative fluids
which include the terms up to the second-order of the dissipative
quantities were developed by M\"{u}ller (non-relativistic)
\cite{Mul67} and Israel and Stewart (relativistic) \cite{Isr76}.
These second-order theories can avoid the problem that the
first-order theories may not be satisfied causality sometimes.  So
they are also called the causal theories of dissipative fluids
dynamics.

Although the Israel-Stewart second-order formalism for relativistic
dissipative fluids \cite{Isr76} was established thirty years ago (in
the 1970s), its numerical implementation in high energy heavy ion
collisions begins in the 21st century.  In Ref. \cite{Mur024}, the viscous
hydrodynamics based on the second-order theory is first used to
study the system expansion in relativistic heavy ion collisions with
the Bjorken scaling hypothesis \cite{Bjo83}.  Recently, the causal
dissipative fluid dynamics has been used in high heavy ion
collisions for the investigations of the effects of shear viscosity
\cite{{Bai06,Cha06a,Cha06,Rom07,Rom07L,Son08L,Son08,Son08a,Luz08,Luz09,
Cha09}} and both shear and bulk viscosities
\cite{Mon09,Son09,Boz09,Li09} on the transverse momentum
distribution \cite{{Bai06,Cha06a,Cha06,Rom07,Son08,Cha09,Mon09,
Son09, Boz09,Li09}}, elliptic flow \cite{{Cha06,Rom07L,Son08L,Son08,
Son08a,Luz08,Luz09,Cha09,Mon09,Son09,Boz09}}, and
Hanbury-Brown-Twiss (HBT) interferometry \cite{Rom07,Boz09,Li09}.

As is well known, a dissipative fluid has not only shear viscosity
but also bulk viscosity.  Recent researches \cite{Mon09,Son09}
indicate that the effects of the shear and bulk viscosities on the
elliptic flow are completely different in the RHIC heavy ion
collisions.  The shear viscosity suppresses the $v_2$ at larger
transverse momentum but the bulk viscosity increases it
\cite{Mon09,Son09}.  Unlike elliptic flow, which reflects the
anisotropic pressure of the system, HBT correlations are related to
the space-time structure of the particle-emitting source.  The
investigations of the shear and bulk viscosity effects on HBT
interferometry may probably provide the messages of the shear and
bulk viscosities from another aspect.

In the present work, we begin with the equations based on the
Israel-Stewart theory for the net baryon free system formed in
ultrarelativistic heavy ion collisions.  For the central collisions,
it is assumed that the system has the Bjorken cylinder geometry
(expanding isotropically in transverse and satisfying longitudinal
boost-invariance) \cite{Bay83,Ris96,Efa05} and undergoes the
crossover transition from the QGP to hadronic gas.  We shall take
into account both the shear and bulk viscosities and solve the
viscous hydrodynamic equations by the relativistic
Harten-Lax-Leer-Einfeldt (RHLLE) algorithm
\cite{RHLLE,Ris9598,Zha04,Efa05}.  By applying a simulated analysis
of two-pion HBT interferometry \cite{Zha9300,Zha04} to the
(1+1)-dimensional viscous hydrodynamic sources, we investigate the
effects of the shear and bulk viscosities on the HBT radii $R_{\rm
out}$, $R_{\rm side}$, and $R_{\rm long}$ \cite{Ber88,Pra90}.  The
HBT analyses indicate that the viscosity effect on the HBT results
is small if one considers only the shear viscosity.  The bulk
viscosity leads to a larger transverse freeze-out configuration of
the pion sources, and thus increases the transverse HBT radii. The
results of the longitudinal HBT radius for the source with Bjorken
longitudinal scaling are consistent with the experimental data.

This paper is organized as follows.  In Sec. \ref{sec:Evo-Equation},
we review briefly the evolution equations of the relativistic
dissipative fluid dynamics for the net baryon free system with the
Bjorken cylinder geometry.  We present the expressions of the
relaxation equations for the shear tensor and bulk pressure for the
system considered.  The detailed derivations are given in Appendix
A.  In Sec. \ref{sec:numeric}, we discuss the numerical algorithm
for solving the viscous hydrodynamics.  The equations of state and
initial conditions used in our calculations are outlined.  We
present our numerical results of the viscous and ideal fluid
evolutions.  In Sec. \ref{sec:HBT}, we perform the two-pion HBT
interferometry for the viscous hydrodynamic sources.  We investigate
the viscosity effects on the HBT radii for the viscous hydrodynamic
sources with only shear viscosity and with both shear and bulk
viscosities.  We also investigate the HBT results for the viscous
sources with lower and higher freeze-out temperatures and initial
energy densities.  Finally, a summary and discussions are given in
Sec. \ref{sec:summery}.

\section{Equations of relativistic dissipative fluid dynamics}
\label{sec:Evo-Equation}

First, we briefly review the evolution equations of relativistic
dissipative fluid dynamics for the system with the Bjorken-cylinder
geometry and zero net baryon density.  The basic equations can be
referred to \cite{MurRis04,Hei06,Bai06,Mur07,Son09t}.

The evolution equations of hydrodynamics come from physics
conservations.  For the baryon free system, we consider only the
energy-momentum conservation in the hydrodynamics.  It is convenient
to describe the Bjorken-cylinder system with the coordinates
$(\tau,\rho,\phi,\eta)$, where $\tau = \sqrt{t^2{-}z^2}$ is the
longitudinal proper time, $\rho$ and $\phi$ are the polar coordinates in
the plane transverse to the beam direction $z$, and $\eta =
\frac{1}{2}\ln[(t{+}z)/(t{-}z)]$ is the space-time rapidity.  The metric
tensors $g^{\mu\nu}$ and $g_{\mu\nu}$ are expressed in the frame of
these curvilinear space-time coordinates as
\begin{eqnarray}
g^{\mu\nu}&=&{\rm
diag}\,\bigl(1,-1,-1/\rho^2,-1/\tau^2\bigr), \nonumber\\
g_{\mu\nu}&=&{\rm diag}\,\bigl(1,-1,-\rho^2,-\tau^2\bigr).
\end{eqnarray}
Some notations used in this paper are
\begin{equation}
d_\mu u^\nu \equiv \partial_\mu u^\nu+\Gamma_{\alpha \mu}^{\nu}
u^\alpha,
\end{equation}
\begin{equation}
\Gamma^{\gamma}_{\alpha \beta} \equiv \frac{1}{2}g^{\gamma \sigma}
\bigl (\partial_{\alpha} g_{\beta \sigma} + \partial_{\beta}
g_{\sigma \alpha} -\partial_{\sigma} g_{\alpha \beta}\bigr),
\end{equation}
\begin{equation}
\Theta \equiv d_\mu u^\mu, ~~~~~~ D \equiv u^\mu d_\mu,
\end{equation}
\begin{equation}
\nabla^\mu \equiv \Delta^{\mu \nu} d_\nu,
\end{equation}
\begin{equation}
\Delta^{\mu\nu} \equiv g^{\mu \nu}-u^\mu u^\nu,
\end{equation}
\begin{equation}
\nabla^{<\mu}u^{\nu>} \equiv \nabla^{(\mu}u^{\nu)}-\frac{1}{3}
\Delta^{\mu\nu}d_\lambda u^{\lambda},
\end{equation}
\begin{equation}
A^{(\alpha\beta)} \equiv (A^{\alpha\beta}+A^{\beta\alpha})/2,
\end{equation}
were $d_\mu$ denotes the covariant derivative, $\partial_\alpha
=\partial /\partial x^\alpha$ and $u^\mu=\gamma (1,\,\vv)$ are the
4-derivative and 4-velocity ($\gamma=1 /\sqrt{1-\vv^2}$),
$\Gamma^{\gamma}_{\alpha \beta}$ is the Christoffel symbol, and
$A^{(\alpha\beta)}$ denotes the symmetrization of the quantity
$A^{\alpha\beta}$.

We adopt the Landau and Lifshitz frame \cite{Lan59} in which the
local energy flow is zero.  The energy-momentum tensor of a fluid
cell in relativistic dissipative hydrodynamics is
\begin{eqnarray}
\label{Tmunu}
T^{\mu\nu} = \varepsilon
u^{\mu}u^{\nu}-(p+\Pi)\Delta^{\mu\nu}+\pi^{\mu\nu},
\end{eqnarray}
where, $\varepsilon$ is the energy density, $p$ is the local
isotropic pressure, $\Pi$ is the bulk viscosity pressure, and
$\pi^{\mu\nu}$ is the shear stress tensor.  Using the covariant
derivative form, the energy-momentum conservation can be reexpressed
as
\begin{eqnarray}
\label{ConsTmn}
d_{\mu}T^{\mu\nu} \equiv \partial_{\mu}T^{\mu \nu}+
\Gamma^{\mu}_{\mu \lambda}T^{\lambda \nu}+T^{\mu
\lambda}\Gamma^{\nu}_{\lambda \mu}= 0 .
\end{eqnarray}

For the Bjorken cylinder, the longitudinal velocity is given by
$v^z=z/t$ and the hydrodynamical solution at an arbitrary
longitudinal coordinate $z$ can be obtained by the Lorentz boost
with the rapidity $\eta=\tanh^{-1}(z/t)$ from the system's
transverse evolution at $z=0$ \cite{Bay83,Ris96,Efa05,Zha09}.  So we
need only to solve the hydrodynamical equations at $z=0$ ($\eta=0$)
in this case.  Considering $u^{\eta}=0$ at $z=0$ and assuming
$u^{\phi}=0$, Eq. (\ref{ConsTmn}) becomes
\begin{eqnarray}
\label{Tmn1}
\left\{
\begin{array}{ll}
\partial_{\tau}T^{\tau\tau}+\partial_{\rho}T^{\rho\tau}+
\Gamma^{\tau}_{\eta\eta}T^{\eta\eta}
+\Gamma^{\eta}_{\eta\tau}T^{\tau\tau}+\Gamma^{\phi}_{\phi
\rho}T^{\rho\tau} = 0, \\[1ex]
\partial_{\tau}T^{\tau \rho}+\partial_{\rho}T^{\rho\rho}+
\Gamma^{\rho}_{\phi\phi}T^{\phi\phi}+\Gamma^{\phi}_{\phi
\rho}T^{\rho\rho}+\Gamma^{\eta}_{\eta\tau}T^{\tau\rho}=0 .
\end{array} \right.
\end{eqnarray}
Introducing the quantities
\begin{equation}
\mathcal {P}_{\rho} = p +\Pi +\pi^{\rho\rho}/\gamma^2,
\end{equation}
\begin{equation}
\mathcal {P}_{\phi} = p +\Pi +\pi^{\phi\phi} \rho^2,
\end{equation}
\begin{equation}
\mathcal {P}_{\eta} = p +\Pi +\pi^{\eta\eta} \tau^2,
\end{equation}
\begin{equation}
E = (\varepsilon+\mathcal {P}_{\rho})\gamma^{2}-\mathcal {P}_{\rho},
\end{equation}
\begin{equation}
M = (E+\mathcal {P}_{\rho}) v_{\rho},
\end{equation}
the non zero components of the energy-momentum tensor $ T^{\mu\nu}$
can be expressed as
\begin{equation}
T^{\tau\tau} = E, ~~~~ T^{\tau\rho}=T^{\rho\tau} = M,
\end{equation}
\begin{equation}
T^{\rho\rho} = M v_{\rho}+\mathcal {P}_{\rho},
\end{equation}
\begin{equation}
T^{\phi\phi} = \frac{\mathcal {P}_{\phi}}{\rho^2},~~~~~~
T^{\eta\eta} = \frac{\mathcal {P}_{\eta}}{\tau^2},
\end{equation}
and, Eq.(\ref{Tmn1}) can be rewritten as
\begin{equation}
\label{Tmn21}
\partial_{\tau}E +\partial_{\rho} [(E+\mathcal {P}_{\rho})\, v_\rho]
=-(E+\mathcal {P}_{\rho})\Big(\frac{v_\rho}{\rho} +
\frac{1}{\tau}\Big) +\frac{(\mathcal {P}_{\rho}-\mathcal
{P}_{\eta})}{\tau},
\end{equation}
\begin{equation}
\label{Tmn22}
\partial_{\tau}M+\partial_{\rho}(M v_\rho +\mathcal {P}_{\rho}) =
-M \Big(\frac{v_\rho}{\rho}+\frac{1}{\tau}\Big) -\frac{(\mathcal
{P}_{\rho}-\mathcal {P}_{\phi})}{\rho}.
\end{equation}
The coupled equations (\ref{Tmn21}) and (\ref{Tmn22}) are the
evolution equations of the dissipative hydrodynamics for the Bjorken
cylinder system, which can be numerically solved by the RHLLE
algorithms \cite{RHLLE,Ris9598,Zha04,Efa05}.  The quantities
$\mathcal {P}_{\rho}$, $\mathcal {P}_{\phi}$, and $\mathcal
{P}_{\eta}$ include the effects of dissipation and they reduce to
the isotropic pressure $p$ for perfect fluids.

Next, we present the relaxation equations that the dissipation
quantities ($\pi^{\mu\nu}$ and $\Pi$) are satisfied for our system.  The
detailed derivations of the relaxation equations are given in
Appendix A. In order to obtain the system evolution we need to solve
the equations (\ref{Tmn21}) and (\ref{Tmn22}) together with the
relaxation equations.

In the Landau and Lifshitz frame and for the baryon free system, the
dissipative effect from the heat conduction can be neglected
\cite{MurRis04,Hei06,Cha06,Son08L,Son08}.  On the other hand, one
can also ignore the effect of vorticity because it is small for the
longitudinally boost-invariant system \cite{Rom07L,Son08a,Luz08}. In
this case, the relaxation equations for the shear tensor
$\pi^{\mu\nu}$ and bulk pressure $\Pi$ can be written as
\cite{MurRis04,Hei06,Bai06,Mur07,Son09t}
\begin{eqnarray}
\label{Trans1} \tau_\pi\Delta^{\mu\alpha} \Delta^{\nu\beta}
D{\pi}_{\alpha\beta}+\pi^{\mu\nu}= 2
\,\widetilde{\eta}\,\nabla^{<\mu}u^{\nu>}-\frac{1}{2}\pi^{\mu\nu}
\widetilde{\eta}\, T d_{\lambda} \Big( \frac{\tau_{\pi}
u^{\lambda}}{\widetilde{\eta}\, T}\Big) ,
\end{eqnarray}
\begin{eqnarray}
\label{Trans2} \tau_\Pi D{\Pi}+\Pi = -\widetilde{\zeta}\, d_\mu
u^\mu -\frac{1}{2}\Pi\, \widetilde{\zeta}\, T d_\lambda \Big(
\frac{\tau_\Pi u^\lambda}{\widetilde{\zeta}\, T} \Big) ,
\end{eqnarray}
where , $\widetilde{\eta}$ and $\widetilde{\zeta}$ denote the shear
and bulk viscous coefficients, $\tau_{\pi}=2\beta_2\widetilde{\eta}$
and $\tau_{\Pi}=\beta_0 \widetilde{\zeta}$ are the corresponding
relaxation times, $\beta_2$ and $\beta_0$ are the relaxation
coefficients for the expansion of entropy flow up to the
second-order of the dissipative quantities, and $T$ is the
temperature.

For the symmetric Bjorken cylinder and at $z=0$ ($u^\phi=u^\eta=0$),
Eqs. (\ref{Trans1}) and (\ref{Trans2}) can be written in the forms
which are suitable for solving numerically as (for detailed derivations,
see Appendix A)
\begin{eqnarray}
\frac{\partial}{\partial \tau} \tau^{\rho\rho}
+\,v_\rho \frac{\partial \tau^{\rho\rho}}{\partial \rho}=
\frac{1}{\gamma\tau_\pi} (-\tau^{\rho\rho} +2\,
\widetilde{\eta}\,\sigma^{\rho\rho}) - \frac{\tau^{\rho\rho}}{2\gamma}
\Big[ \Theta +\frac{\widetilde{\eta}\,T}{\tau_\pi} D \Big(
\frac{\tau_\pi}{\widetilde{\eta}\,T} \Big) \Big], \label{T.rr}
\end{eqnarray}
\begin{eqnarray}
\frac{\partial}{\partial\tau} \tau^{\phi\phi}
+\,v_\rho \frac{\partial\tau^{\phi\phi}}{\partial \rho}
= \frac{1}{\gamma \tau_\pi} (-\tau^{\phi\phi} +2\,
\widetilde{\eta}\,\sigma^{\phi\phi}) - \frac{\tau^{\phi\phi}}{2\gamma}
\Big[ \Theta +\frac{\widetilde{\eta}\,T}{\tau_\pi} D \Big(
\frac{\tau_\pi}{\widetilde{\eta}\,T} \Big) \Big], \label{T.phi}
\end{eqnarray}
\begin{eqnarray}
\frac{\partial}{\partial\tau}\tau^{\eta\eta}
+\,v_\rho \frac{\partial \tau^{\eta\eta}}{\partial \rho}
= \frac{1}{\gamma \tau_\pi}(-\tau^{\eta\eta} + 2\,
\widetilde{\eta}\, \sigma^{\eta\eta}) -\frac{\tau^{\eta\eta}}{2\gamma}
\Big[ \Theta + \frac{\widetilde{\eta}\,T}{\tau_\pi} D \Big(
\frac{\tau_\pi}{\widetilde{\eta}\,T} \Big) \Big], \label{T.eta}
\end{eqnarray}
\begin{eqnarray}
\frac{\partial}{\partial\tau} \Pi+\,v_\rho \frac{\partial\Pi}{\partial
\rho} = \frac{1}{\gamma \tau_\Pi} (-\Pi
-\widetilde{\zeta}\,\Theta) -\frac{\Pi}{2\gamma} \Big[\Theta
+\frac{\widetilde{\zeta}\, T}{\tau_\Pi} D \Big (\frac{\tau_
\Pi}{\widetilde{\zeta}\,T}\Big) \Big], \label{T.Pi}
\end{eqnarray}
where $\tau^{\rho\rho}$, $\tau^{\phi\phi}$, and $\tau^{\eta\eta}$
are three introduced quantities whith the relations to the
nonzero components of $\pi^{\mu\nu}$ as
\begin{eqnarray}
\pi^{\mu\nu} = \left(
\begin{array}{cccc}
\gamma^2 v_\rho^2\tau^{\rho\rho} & \gamma^2 v_\rho \tau^{\rho\rho}& 0 &0 \\
\gamma^2 v_\rho\tau^{\rho\rho}& \gamma^2 \tau^{\rho\rho} & 0 & 0 \\
0 &0 & \rho^{-2} \tau^{\phi\phi} &0 \\
0&  0 & 0 & \tau^{-2} \tau^{\eta\eta}
\end{array}
\right)\,,
\end{eqnarray}
where
\begin{equation}
\tau^{\rho\rho}+\tau^{\phi\phi}+\tau^{\eta\eta}=0, \label{ttt}
\end{equation}
from the traceless condition $\pi_\mu^\mu=0$ \cite{Mur024,Bai06}.
In Eqs. (\ref{T.rr}) -- (\ref{T.Pi}),
\begin{eqnarray}
\Theta = d_{\mu} u^{\mu} = \frac{\partial\gamma}{\partial \tau}
+\frac{\partial (\gamma\,v_\rho)}{\partial \rho} + \gamma
\Big(\frac{v_{\rho}}{\rho} + \frac{1}{\tau}\Big)
\equiv \theta + \gamma \Big(\frac{v_{\rho}}{\rho} + \frac{1}{\tau}\Big),
\end{eqnarray}
\begin{equation}
\sigma^{\rho\rho}=\Big(-\theta + \frac{1}{3}\Theta\Big),
\end{equation}
\begin{equation}
\sigma^{\phi\phi} = \Big(\frac{-\gamma\, v_\rho}{\rho} +
\frac{1}{3}\Theta\Big),
\end{equation}
\begin{equation}
\sigma^{\eta\eta} = \Big(\frac{-\gamma}{\tau} +
\frac{1}{3}\Theta\Big),
\end{equation}
and $D=u^\mu d_\mu$ reduce to $\gamma (\partial_\tau +v_\rho
\partial_\rho)$.

When performing numerical calculations we need to solve the evolution
equations (\ref{Tmn21}) and (\ref{Tmn22}) and the relaxation
equations (\ref{T.rr}) -- (\ref{T.Pi}) simultaneously.  For the
sQGP, the ratio of the shear viscosity to the entropy density
satisfies $\widetilde{\eta}/s \,\undersim > 1/4\pi$ \cite{KSS03,BucLiu04,KSS05},
and the ratio of the bulk to shear viscosities is proportional to the deviation
of the sound velocity square from that of the ideal hadronic gas,
$\widetilde{\zeta}/\widetilde{\eta} \simeq -\kappa (v_s^2
-1/3)$ \cite{BenBuc05,Buc05,Buc08,Son09}.  Based on the calculations of the
${\cal N}=2^*$ gauge theory, the value of $\kappa$ is between 3.142
and 4.935 \cite{Buc08}.  In our calculations we take $\widetilde{\eta}/s
=2\,C_s (1/4\pi)$ $(C_s=1,2)$ and $\kappa=4.75$.  The relaxation times
for the shear and bulk viscosities are taken as $\tau_{\pi}=\frac{6{\widetilde
\eta}}{sT}$ and $\tau_{\Pi}= \tau_{\pi}$ as in
Refs. \cite{Isr76,Rom07,Son08L,Son08a,Cha09,Son09,Boz09} for simplicity.

\section{Numerical solution of viscous hydrodynamics}
\label{sec:numeric}

When solving the viscous hydrodynamic equations numerically, we
encounter the following two types of equations:
\begin{eqnarray}
\partial_{\tau} U + \partial_{\rho} F(U)=G(U)\label{UE1}
\end{eqnarray}
and
\vspace*{-10mm}
\begin{eqnarray}
\partial_{\tau} U + v_{\rho} \partial_{\rho} F(U)=G(U). \label{UE2}
\end{eqnarray}
For the evolution equations (\ref{Tmn21}) and (\ref{Tmn22}), $U$
in Eq. (\ref{UE1}) can be replaced by $E$ or $M$, and for the
relaxation equations (\ref{T.rr}) -- (\ref{T.Pi}), U in Eq.
(\ref{UE2}) can be replaced by $\tau^{\rho\rho}$, $\tau^{\phi\phi}$,
$\tau^{\eta\eta}$, or $\Pi$. Eq. (\ref{UE1}) can be solved directly
by the RHLLE algorithm \cite{RHLLE,Ris9598,Zha04,Efa05}.  For Eq.
(\ref{UE2}) we can change it to the type of Eq. (\ref{UE1}) by
adding the term $F(U)\,\partial_{\rho} v_{\rho}$ to both sides of the
equation, and then solve it with the RHLLE algorithm.  In the
calculations, we solve simultaneously the six coupled equations of
(\ref{Tmn21}), (\ref{Tmn22}), and (\ref{T.rr}) -- (\ref{T.Pi}) for
the quantities $(E, M, \tau^{rr}, \tau^{\phi\phi},
\tau^{\eta\eta},\Pi)$ at each time step.  The width of time step is
taken to be $\Delta \tau=0.04$ fm/c and the width of space step is
taken to be $0.99\Delta \tau$ fm.  The set of the coupled equations
(\ref{Tmn21}), (\ref{Tmn22}), and (\ref{T.rr}) -- (\ref{T.Pi}) is
closed by the equation of state (EOS), and the transverse velocity
and local energy density satisfy
\begin{eqnarray}
v_\rho=\frac{M}{E+P_\rho},\hspace{1em} \varepsilon=E -
\frac{M^2}{E+P_\rho}.\hspace{1em}
\end{eqnarray}

\begin{figure}
\includegraphics*[width=6.0cm]{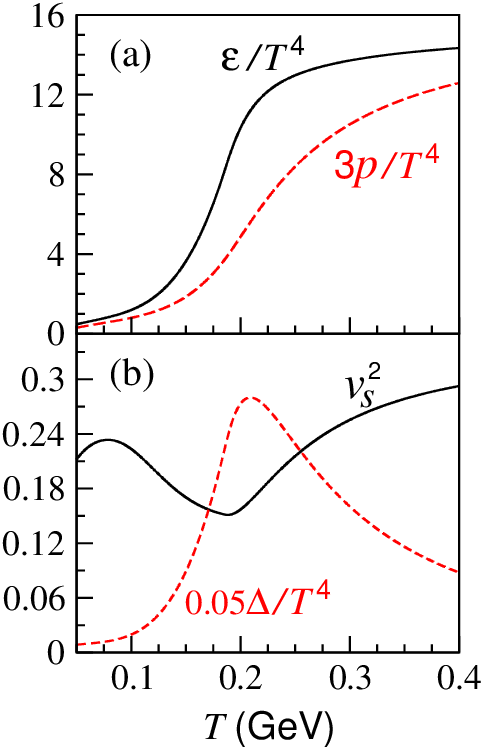}
\caption{(Color online) (a) The energy density $\varepsilon$ and pressure
$p$ as functions of temperature.  (b) The square of sound velocity $v_s^2$
and the difference $\Delta=\varepsilon-3p$ as functions of temperature.}
\label{EOS}
\end{figure}

The EOS used in the calculations is the parametric EOS which
combines hadron resonance gas at low temperatures with lattice QCD
at high temperatures \cite{Huo10}.  Figure \ref{EOS} (a) shows the
energy density $\varepsilon$ and pressure $p$ as functions of
temperature. Figure \ref{EOS} (b) shows the square of sound velocity
$v_s^2$ and the difference between $\varepsilon$ and $3p$,
$\Delta=\varepsilon -3p$, as functions of temperature.  The
transition temperature $T_c$ is taken to be 170 MeV.

\begin{figure}
\includegraphics*[width=6.5cm]{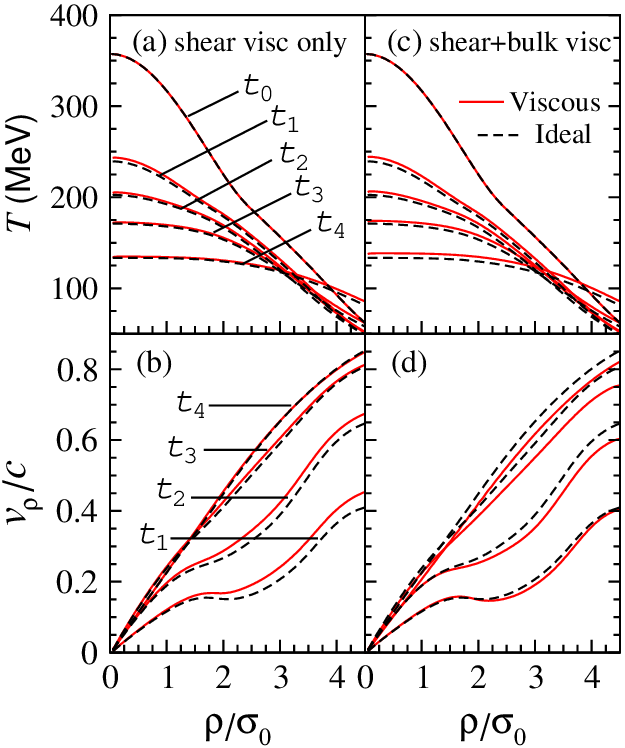}
\caption{(Color online) The temperatures and transverse velocities of the
viscous fluids with only shear viscosity ($C_s=2$) and with the shear and
bulk viscosities, in $z=0$ plane.  The labeled time is respectively $t_0=\tau_0$,
$t_n =\tau_0 + 2^n (20\Delta \tau)$ ($n=1,2,3,4,~\Delta \tau=0.04$ fm/c).}
\label{Ftv}
\end{figure}

On the basis of the Bjorken picture \cite{Bjo83}, the system evolves
hydrodynamically after a proper time $\tau_0$.  In our calculations,
we take the initial time $\tau_0 =0.6$ fm/c, and use a Gaussian
initial transverse distribution of energy density at $z=0$,
\begin{equation}
\label{IE} \varepsilon (\rho, z=0)= \varepsilon_0
\exp(-\rho^2/2\sigma_0),
\end{equation}
where $\varepsilon_0$ is taken to be 30 GeV/fm$^3$ and $\sigma_0$ is
taken to be 3.3 fm.  The initial shear tensor components are taken
to be $\tau^{\rho\rho}(\rho) = \tau^{\phi\phi}(\rho)=
\frac{\varepsilon(\rho)}{18}$ and $\tau^{\eta\eta}(\rho) =
-\frac{\varepsilon(\rho)}{9}$ as in Ref. \cite{MurRis04}.  We find
that the system evolution is almost independent of the initial value
of the bulk pressure because it is almost zero at the initial
temperature much higher than $T_c$.  So we take the initial bulk
pressure to be zero in the calculations.

Figures \ref{Ftv} (a) and (b) show the temperature and transverse
velocity profiles in $z=0$ plane for the viscous fluid with only
shear viscosity ($C_s=2$).  Correspondingly, Figs. \ref{Ftv} (c) and
(d) show the temperature and transverse velocity profiles in $z=0$
plane for the fluid with the shear and bulk viscosities.  In Fig.
\ref{Ftv}, the dashed lines are for the ideal fluid for comparison,
and the time $t_0=\tau_0$, $t_n =\tau_0 + 2^n (20\Delta \tau)$
($n=1,2,3,4$, $\Delta \tau=0.04$ fm/c).  For the only shear
viscosity case the viscous fluids cool a little more slowly than
the ideal fluid for small time, and the transverse velocities of the
viscous fluid are higher than the corresponding velocities of ideal
fluid.  Once the bulk viscosity is taken into account, one can see
that the viscous fluid cools more slowly and its transverse
velocities are lower than the corresponding velocities of ideal
fluid.  This is because that the bulk viscosity decreases the system
pressure (see Eq. (\ref{Tmunu}) and notice $\Pi<0$) and therefore
decreases the gradient of the pressures in and out of the system.

\begin{figure}
\includegraphics*[width=7cm]{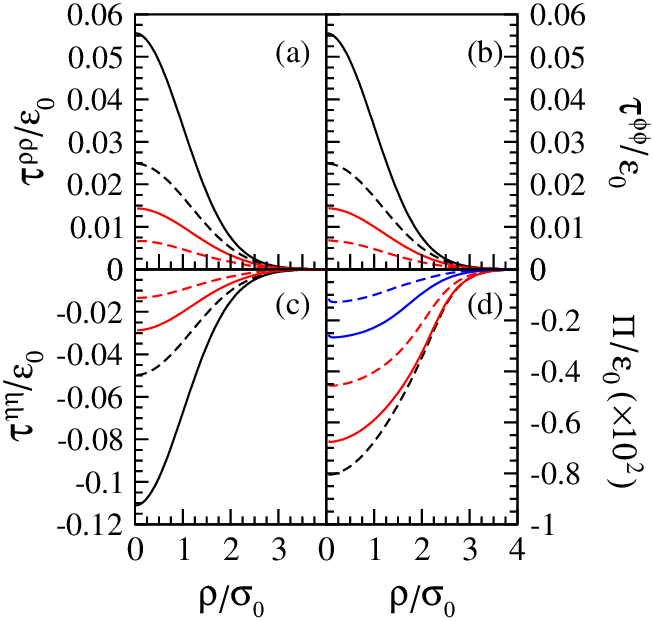}
\caption{(Color online) The ratios of $\tau^{\rho\rho}$,
$\tau^{\phi\phi}$, $\tau^{\eta\eta}$, and $\Pi$ to $\varepsilon_0$ as
functions of the transverse coordinate and time for the viscous fluid
($C_s=2$).  Here, the time is taken to be $\tau_0$
(solid black), $\tau_0+6\Delta \tau$ (dashed black), $\tau_0+12\Delta
\tau$ (solid red), $\tau_0+24\Delta \tau$ (dashed red), $\tau_0+
48\Delta \tau$ (solid blue), and $\tau_0+96\Delta \tau$ (dashed blue),
respectively. ($\Delta \tau=0.04$ fm/c) }
\label{Ftau}
\end{figure}

\begin{figure}
\includegraphics*[width=6.5cm]{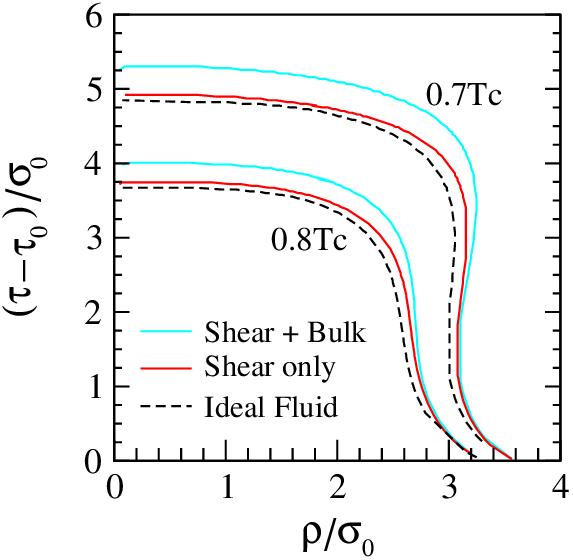}
\caption{(Color online) The isotherms for 0.7 and 0.8$\,T_c$ in $z=0$
plane.} \label{Ftfreeze}
\end{figure}

In Fig. \ref{Ftau}, we show the ratios of the viscosity quantities
$\tau^{\rho\rho}$, $\tau^{\phi\phi}$, $\tau^{\eta\eta}$, and $\Pi$
to the central initial energy density $\varepsilon_0$, as functions
of the transverse coordinate and time for the viscous fluid
($C_s=2$).  In Fig.\ref{Ftfreeze}, we draw the isotherms for the
viscous ($C_s=2$) and ideal fluids in $z=0$ plane.  It can be seen
that the contours for the only shear viscous fluid are slightly larger
than that of the ideal fluid, and the bulk viscosity lets the
contours of the viscous fluid be much larger than those of the ideal
fluid.  This is consistent with the results in Fig. \ref{Ftv}.

After knowing the transverse evolution at $z=0~(t=\tau)$, we can
obtain the temperature and velocity at arbitrary $z$ for the Bjorken
cylinder system by \cite{Bay83,Ris96,Efa05,Zha09}
\begin{equation}
T(t, \rho, z)=T(\tau, \rho, 0),
\end{equation}
\begin{equation}
v_{\rho}(t, \rho, z)=v_{\rho}(\tau, \rho, 0) \frac{\tau}{t}, ~~~~
v_z=\frac{z}{t}.
\end{equation}
Then, we can calculate the two-pion HBT correlation functions by a
simulation \cite{Zha9300,Zha04}.

\section{Pion interferometry analysis}
\label{sec:HBT}

The two-pion HBT correlation function is defined as the ratio of the
two-pion momentum distribution $P(k_1,k_2)$ to the the product of
the single-pion momentum distribution $P(k_1) P(k_2)$.  For a
chaotic pion-emitting source, $P(k_i)~(i=1,2)$, and $P(k_1,k_2)$ can
be expressed as \cite{Won94}
\begin{eqnarray}
\label{Pp1}
P(k_i) = \sum_{X_i} A^2(k_i,X_i) \,,
\end{eqnarray}
\begin{eqnarray}
\label{Pp12}
P(k_1,k_2) = \sum_{X_1, X_2} \Big|\Phi(k_1, k_2; X_1,
X_2 )\Big|^2 ,
\end{eqnarray}
where $A(k_i,X_i)$ is the magnitude of the amplitude for emitting a
pion with 4-momentum $k_i=({\bf k}_i,E_i)$ at $X_i$ and $\Phi(k_1,
k_2; X_1, X_2 )$ is the two-pion wave function. Assuming that the
emitted pions propagate as free particles, $\Phi(k_1, k_2; X_1, X_2
)$ is simply
\begin{eqnarray}
\label{PHI} & &\!\!\!\!\!\!\!
\Phi(k_1, k_2; X_1, X_2 )
\nonumber \\
& & =\frac{1}{\sqrt{2}} \Big[ A(k_1, X_1) A(k_2, X_2) e^{i k_1 \cdot
X_1 + i k_2 \cdot X_2}
\nonumber \\
& & ~~~+ A(k_1, X_2) A(k_2, X_1) e^{i k_1 \cdot X_2 + i k_2 \cdot
X_1 } \Big] .~~~~~
\end{eqnarray}

For a set of variables of the relative momentum, $\{q_j\}$
($q_j=|{\bf k_1}-{\bf k_2}|_j$), the two-pion correlation function
can be expressed as \cite{Zha9300}
\begin{equation}
\label{Cqj}
C(\{q_j\})=\frac{{\rm Cor}(\{q_j\})}{{\rm
Uncor}(\{q_j\})},
\end{equation}
where
\begin{equation}
\label{Cor}
{\rm Cor}(\{q_j\})= \int d{\bf k}_1 d{\bf k}_2
P(k_1,k_2) \prod_j \delta (|{\bf k}_1 - {\bf k}_2|_j -q_j)
\end{equation}
and
\begin{equation}
\label{Uncor}
{\rm Uncor}(\{q_j\})= \int d{\bf k}_1 d{\bf k}_2
P(k_1)P(k_2) \prod_j \delta (|{\bf k}_1 - {\bf k}_2|_j -q_j)
\end{equation}
are the correlated and uncorrelated pion pair distributions with
$\{q_j\}$.

In our simulated calculations, we first generate the pion momentum
$k$ on the freeze-out hypersurface $\Sigma (X)$ with temperature
$T_f$, according to the probability of the Cooper-Frye formula
\cite{CF74}
\begin{equation}
\label{CooF}
P(k) \propto \int k^\mu d \Sigma_\mu f \Big(\frac{k_\mu u^\mu}{T}\Big).
\end{equation}
For the viscous fluid, the distribution $f=f_0 + \delta f$. We take
$f_0$ for the ideal fluid as the Boltzmann distribution, and take
$\delta f$ for the shear and bulk viscosities as \cite{Dus08}
\begin{equation}
\delta f = f_0 \Big(\frac{k_\mu u^\mu}{T}\Big) \frac{k^\mu
k^\nu}{2(\varepsilon + p) T^2} \Big (\pi_{\mu\nu} - \frac{2}{5}
\Pi \Delta_{\mu\nu} \Big).
\end{equation}
In Cartesian frame, $\pi^{\mu\nu}$ can be expressed in terms of the
cylindrical variables as \cite{MurRis04}
\begin{eqnarray}
\pi^{\mu\nu} = \left(
\begin{array}{cccc}
\Pi^{\rho\eta}\cosh^2\!\eta-\tau^{\eta\eta} &
\tau^{\rho\rho}\gamma^2 v_\rho \cos\phi \cosh\eta &
\tau^{\rho\rho}\gamma^2 v_{\rho} \sin\phi \cosh\eta &
\Pi^{\rho\eta} \cosh\eta \sinh\eta \\
\tau^{\rho\rho} \gamma^2 v_\rho \cos\phi \cosh\eta &
\Pi^{\rho\phi} \cos^2\!\phi + \tau^{\phi\phi} &
\Pi^{\rho\phi} \cos\phi \sin\phi &
\tau^{\rho\rho}\gamma^2 v_\rho \cos\phi \sinh\eta  \\
\tau^{\rho\rho} \gamma^2 v_\rho \sin\phi \cosh\eta &
\Pi^{\rho\phi} \cos\phi \sin\phi &
\Pi^{\rho\phi} \sin^2\!\phi + \tau^{\phi\phi} &
\tau^{\rho\rho} \gamma^2 v_\rho \sin\phi \sinh\eta \\
\Pi^{\rho\eta} \cosh\eta \sinh\eta &
\tau^{\rho\rho}\gamma^2 v_\rho \cos\phi \sinh\eta &
\tau^{\rho\rho} \gamma^2 v_\rho \sin\phi \sinh\eta &
\Pi^{\rho\eta}\cosh^2\!\eta-\tau^{\rho\rho}\gamma^2 v_{\rho}^2
\end{array}
\right), \hspace*{-10mm} \nonumber\\
\end{eqnarray}
where $\mu, \nu = t, x, y, z$,
$\Pi^{\rho\eta}=\tau^{\rho\rho}\gamma^2 v_{\rho}^2 +
\tau^{\eta\eta}$, and $\Pi^{\rho\phi}=\tau^{\rho\rho} \gamma^2
-\tau^{\phi\phi}$.

After obtaining the momenta $k_i~(i=1,2)$ of the pions emitted on
the freeze-out hypersurface, we can construct the correlation
functions for the variables $q_{\rm out}$, $q_{\rm side}$, and
$q_{\rm long}$ \cite{Ber88,Pra90}, $ C (q_{\rm out},q_{\rm
side},q_{\rm long})$, based on Eqs. (\ref{Pp1}) --- (\ref{Uncor}),
by summing over ${\bf k}_1$ and ${\bf k}_2$ for each $(q_{\rm
out},q_{\rm side},q_{\rm long})$ bin.  Then, we extract the HBT
radii $R_{\rm out}$, $R_{\rm side}$, and $R_{\rm long}$ by fitting
the correlation function with the parametrized formula
\begin{equation}
C(q_{\rm out}, q_{\rm side}, q_{\rm long})=1+\lambda\,e^{-q^2_{\rm
out} R^2_{\rm out} -q^2_{\rm side} R^2_{\rm side} -q^2_{\rm long}
R^2_{\rm long}}.
\end{equation}

\begin{figure}
\includegraphics*[width=7cm]{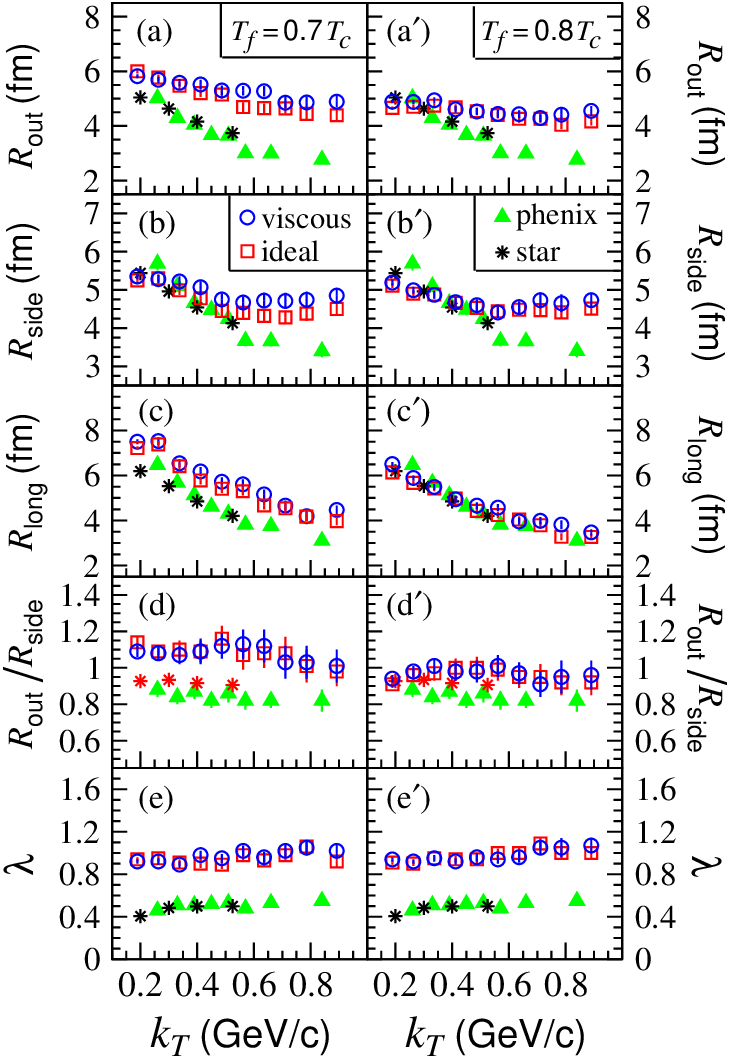}
\caption{(Color online) The two-pion HBT results for the viscous
($C_s=2$) and ideal hydrodynamic sources with $T_f=0.7T_c$
and $T_f=0.8T_c$.  The experimental data are from PHENIX \cite{PHE04}
and STAR \cite{STA05} Collaborations. } \label{Fhbt1}
\end{figure}

Figure \ref{Fhbt1} shows the HBT radii $R_{\rm out}$, $R_{\rm
side}$, $R_{\rm long}$, and $\lambda$ parameter as functions of the
transverse momentum of the pion pair, $k_T=|{\bf k}_{1T} + {\bf k}_{2T}|
/2$, for the viscous and ideal hydrodynamic sources with the freeze-out
temperatures 0.7 and 0.8$T_c$.  For the viscous sources, the
parameter $C_s$ is taken to be 2.  For comparison the RHIC
experimental HBT results \cite{PHE04,STA05} are also shown in the
figure.  It can be seen that the HBT radii for the lower freeze-out
temperature are larger than those for the higher $T_f$.  The
viscosity increases the transverse HBT radii $R_{\rm out}$ and
$R_{\rm side}$ for the the source with the lower freeze-out
temperature.  By comparing with the experimental results, we find
that the results of the longitudinal HBT radius $R_{\rm long}$ for
the hydrodynamic sources with $T_f=0.8T_c$ agree well with the
experimental data.  It indicates that the Bjorken scaling hypothesis
is suitable in the description of the source longitudinal evolution.
In our calculations, the width of the Bjorken rapidity plateau is
taken to be 1.3.  On can also see that the transverse HBT radii
$R_{\rm out}$ and $R_{\rm side}$ as functions of $k_T$ are flat for both
of the two freeze-out temperatures.  However, the ratio $R_{\rm
out}/R_{\rm side}$ decreases to about one for the higher $T_f$.  We
did not consider the source coherence and the Coulomb interaction
between the pions, so the $\lambda$ results of the hydrodynamical
sources are about unit.

\begin{table}
\caption{\label{tabimgr}The HBT results for the ideal and viscous
hydrodynamic sources with $T_f=0.8 T_c$.}
\begin{ruledtabular}
\begin{tabular}{ccccccc}
$k_T\,({\rm GeV/c})$&sources&$R_{\rm out}\,({\rm fm})$&$R_{\rm side}\,({\rm fm})$&
$R_{\rm long}\,({\rm fm})$&$R_{\rm out}/R_{\rm side}$&$\lambda$\\
\hline\\[-4ex]
0.24&ideal fluid&4.69 $\pm$ 0.04&5.01 $\pm$ 0.04&5.89 $\pm$ 0.06&0.94 $\pm$ 0.02&0.91 $\pm$ 0.01\\
   &visc fluid 1&4.72 $\pm$ 0.04&4.95 $\pm$ 0.04&5.96 $\pm$ 0.06&0.95 $\pm$ 0.02&0.92 $\pm$ 0.01\\
   &visc fluid 2&4.80 $\pm$ 0.04&4.99 $\pm$ 0.04&5.98 $\pm$ 0.06&0.96 $\pm$ 0.02&0.93 $\pm$ 0.01\\
   &visc fluid 3&4.82 $\pm$ 0.04&4.96 $\pm$ 0.05&6.00 $\pm$ 0.06&0.97 $\pm$ 0.02&0.92 $\pm$ 0.01\\
   &visc fluid 4&4.89 $\pm$ 0.04&5.02 $\pm$ 0.05&6.16 $\pm$ 0.06&0.97 $\pm$ 0.02&0.93 $\pm$ 0.01\\
\hline\\[-4ex]
0.39&ideal fluid&4.68 $\pm$ 0.06&4.73 $\pm$ 0.07&5.06 $\pm$ 0.08&0.99 $\pm$ 0.03&0.93 $\pm$ 0.02\\
   &visc fluid 1&4.63 $\pm$ 0.06&4.77 $\pm$ 0.07&4.91 $\pm$ 0.08&0.97 $\pm$ 0.03&0.92 $\pm$ 0.02\\
   &visc fluid 2&4.66 $\pm$ 0.06&4.76 $\pm$ 0.06&4.85 $\pm$ 0.07&0.98 $\pm$ 0.03&0.91 $\pm$ 0.02\\
   &visc fluid 3&4.67 $\pm$ 0.06&4.79 $\pm$ 0.07&4.95 $\pm$ 0.08&0.97 $\pm$ 0.03&0.91 $\pm$ 0.02\\
   &visc fluid 4&4.77 $\pm$ 0.07&4.81 $\pm$ 0.07&5.09 $\pm$ 0.08&0.99 $\pm$ 0.03&0.90 $\pm$ 0.02\\
\hline\\[-4ex]
0.55&ideal fluid&4.46 $\pm$ 0.08&4.51 $\pm$ 0.08&4.38 $\pm$ 0.08&0.99 $\pm$ 0.03&0.95 $\pm$ 0.02\\
   &visc fluid 1&4.45 $\pm$ 0.08&4.48 $\pm$ 0.08&4.28 $\pm$ 0.08&0.99 $\pm$ 0.04&0.95 $\pm$ 0.02\\
   &visc fluid 2&4.49 $\pm$ 0.08&4.52 $\pm$ 0.08&4.17 $\pm$ 0.08&0.99 $\pm$ 0.04&0.94 $\pm$ 0.02\\
   &visc fluid 3&4.49 $\pm$ 0.08&4.63 $\pm$ 0.09&4.40 $\pm$ 0.09&0.97 $\pm$ 0.04&0.90 $\pm$ 0.03\\
   &visc fluid 4&4.52 $\pm$ 0.08&4.60 $\pm$ 0.09&4.48 $\pm$ 0.09&0.98 $\pm$ 0.04&0.92 $\pm$ 0.03\\
\hline\\[-4ex]
0.70&ideal fluid&4.25 $\pm$ 0.09&4.51 $\pm$ 0.10&3.88 $\pm$ 0.09&0.94 $\pm$ 0.04&0.98 $\pm$ 0.03\\
   &visc fluid 1&4.23 $\pm$ 0.09&4.56 $\pm$ 0.10&3.92 $\pm$ 0.09&0.93 $\pm$ 0.04&0.99 $\pm$ 0.03\\
   &visc fluid 2&4.23 $\pm$ 0.09&4.59 $\pm$ 0.10&3.78 $\pm$ 0.09&0.92 $\pm$ 0.04&0.95 $\pm$ 0.03\\
   &visc fluid 3&4.26 $\pm$ 0.10&4.56 $\pm$ 0.11&3.91 $\pm$ 0.10&0.93 $\pm$ 0.04&0.95 $\pm$ 0.03\\
   &visc fluid 4&4.30 $\pm$ 0.10&4.67 $\pm$ 0.11&3.89 $\pm$ 0.10&0.92 $\pm$ 0.04&0.94 $\pm$ 0.03\\
\hline\\[-4ex]
0.84&ideal fluid&4.23 $\pm$ 0.12&4.42 $\pm$ 0.13&3.37 $\pm$ 0.10&0.96 $\pm$ 0.05&0.97 $\pm$ 0.04\\
   &visc fluid 1&4.24 $\pm$ 0.12&4.46 $\pm$ 0.12&3.43 $\pm$ 0.10&0.95 $\pm$ 0.05&0.97 $\pm$ 0.04\\
   &visc fluid 2&4.32 $\pm$ 0.12&4.40 $\pm$ 0.12&3.59 $\pm$ 0.10&0.98 $\pm$ 0.05&1.01 $\pm$ 0.04\\
   &visc fluid 3&4.44 $\pm$ 0.13&4.48 $\pm$ 0.13&3.60 $\pm$ 0.11&0.99 $\pm$ 0.06&1.02 $\pm$ 0.04\\
   &visc fluid 4&4.77 $\pm$ 0.14&4.85 $\pm$ 0.15&3.67 $\pm$ 0.12&0.98 $\pm$ 0.06&1.05 $\pm$ 0.05\\
\end{tabular}
\end{ruledtabular}
\end{table}

In Table I we list the HBT results for the ideal and viscous
hydrodynamical sources with $T_f=0.8T_c$.  Here the results of the
`visc fluid 1' and `visc fluid 2' are for the viscous sources with
only shear viscosity for $C_s=1$ and $C_s=2$.  The results of the
`visc fluid 3' and `visc fluid 4' are for the viscous sources with
both the shear and bulk viscosities for $C_s=1$ and $C_s=2$.  One
can see that for the higher freeze-out temperature, the viscosity
effect on the HBT results for the only shear viscous sources are
negligible, and the viscosity effect for the source with both the
shear and bulk viscosities may let the HBT radii be a little larger
than those of the ideal fluid source.

\begin{figure}
\includegraphics*[width=7cm]{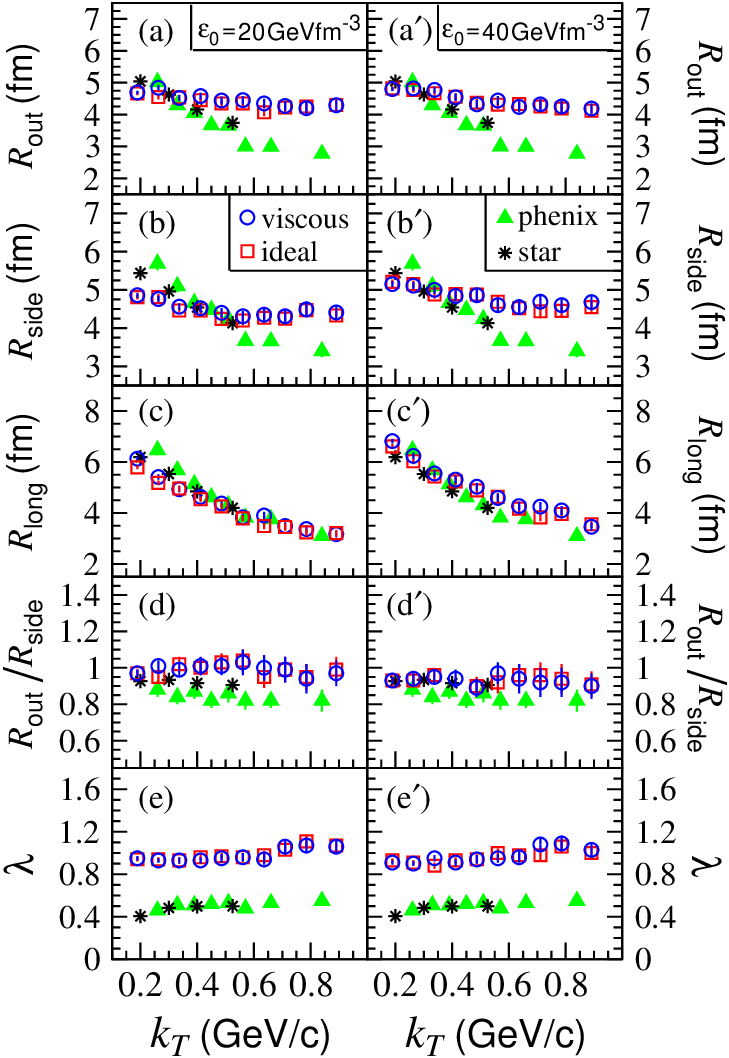}
\caption{(Color online) The two-pion HBT results for the viscous
($C_s=2$) and ideal hydrodynamic sources with $\varepsilon_0
=20$ GeV/fm$^3$ and $\varepsilon_0=40$ GeV/fm$^3$.  The freeze-out
temperature is 0.8$T_c$.  The experimental data are from PHENIX
\cite{PHE04} and STAR \cite{STA05} Collaborations.} \label{Fhbt2}
\end{figure}

Further, we examine the HBT radii for the viscous hydrodynamical
sources with lower initial energy density ($\varepsilon_0=20$
GeV/fm$^3$) and higher initial energy density ($\varepsilon_0=40$
GeV/fm$^3$).  The results are shown in Fig. \ref{Fhbt2}, where the
freeze-out temperature is 0.8$T_c$.  It can be seen that the results
of the transverse HBT radius $R_{\rm side}$ for the higher initial
energy density are larger than those for the lower initial energy
density.  This leads to the smaller values of the ratio $R_{\rm
out}/R_{\rm side}$ for the source with higher $\varepsilon_0$.  The
longitudinal HBT radius $R_{\rm long}$ also increases with the
initial energy density.  However, the transverse HBT radii as
functions of the transverse momentum $k_T$ are more flat than those
of the experimental data, both for the lower and higher initial
energy density sources.

\section{Summery and discussions}
\label{sec:summery}

Based on the Israel-Stewart second-order theory of relativistic
dissipative fluid dynamics \cite{Isr76}, we investigate the
space-time evolution of the viscous hydrodynamics with the Bjorken
cylinder geometry.  The EOS used in the hydrodynamic calculations
combines hadron resonance gas at low temperatures with lattice QCD
at high temperatures \cite{Huo10}.  We consider both the shear and
bulk viscosities in the hydrodynamic evolution of the dissipative
fluids.  By performing the two-pion HBT interferometry to the viscous
and ideal hydrodynamic sources, we investigate the effects of the
viscosities on the two-pion HBT interferometry results.

For a dissipative fluid there is not only shear viscosity but also
bulk viscosity.  For the sQGP produced in relativistic heavy ion
collisions \cite{RHIC}, the ratio of the shear viscosity to the
entropy density satisfies $\widetilde{\eta}/s \,\undersim > 1/4\pi$
\cite{KSS03,BucLiu04,KSS05}, and the bulk viscosity
$\widetilde{\zeta}$ is proportional to $\widetilde{\eta}(v_s^2
-1/3)$ \cite{BenBuc05,Buc05,Buc08,Son09}.  Our calculations indicate
that the viscosity effect on the two-pion HBT results is small if
only the shear viscosity is considered.  The bulk viscosity leads to
a larger transverse freeze-out configuration of the pion sources,
and therefore increases the transverse HBT radii.  This effect is larger
if the source has a lower freeze-out temperature.  In the present
paper we took the relaxation time $\tau_{\pi}=\frac{6{\widetilde \eta}}{sT}$
\cite{Isr76,Rom07,Son08L,Son08a,Cha09} and $\tau_{\Pi}= \tau_{\pi}$
\cite{Son09,Boz09}, and used an uniform formula of the shear
viscosities for both the QGP and hadronic phases for simplicity.  We find 
that there are not observable influences on the viscosity effects on HBT
results for the $\tau_{\pi}$ and $\tau_{\Pi}$ values within
reasonable ranges.  In Ref. \cite{Boz09} different viscosities for
the QGP and hadronic phases are taken into account in the analyses
of transverse momentum spectra, elliptic flow, and HBT
interferometry.  Further investigations of the effects of shear and
bulk viscosities on the HBT interferometry for more realistic
viscous pion-emitting sources in relativistic heavy ion collisions
are of great interest in the future.

In our model the Bjorken longitudinal scaling hypothesis
\cite{Bjo83} is adopted, which is an approximation for the heavy ion
collisions at very high energies.  Our results indicate that the
longitudinal HBT radius for the Bjorken cylinder hydrodynamic source
agrees well with the RHIC HBT results.  However the transverse HBT
radii as functions of the pair transverse momentum for our Bjorken
cylinder sources are more flat than those of the experimental data.
In order to explain the RHIC HBT results, various models without
viscosity were proposed
\cite{Lin02,Soc04,Zha04,Zha06,Cra05,Fro06,Bro08}. Recently, the
hydrodynamic source model including the initial collective flow,
stiffer EOS, and viscosity was proposed \cite{Pra09} to explain the
RHIC HBT results.  The comprehensive examinations of the effects in
these models on HBT interferometry and other observables (e.g.
particle momentum distributions, elliptic flow, etc.) require
considerable efforts.

\begin{acknowledgements}
The authors would like to thank Dr. S. Khakshurnia for helpful
discussions.  This work was supported by the National Natural
Science Foundation of China under Contract No. 11075027.
\end{acknowledgements}


\appendix
\section{Derivation for relaxation equations}
Here we give the derivations of Eqs. (\ref{T.rr}) -- (\ref{T.eta})
from Eq. (\ref{Trans1}) in detail.  For the Bjorken cylinder and at
$z=0$, we have
\begin{eqnarray}
\pi^{\mu\nu} = \left(
\begin{array}{cccc}
\gamma^2 v_\rho^2\tau^{\rho\rho} & \gamma^2 v_\rho \tau^{\rho\rho}& 0 &0 \\
\gamma^2 v_\rho\tau^{\rho\rho}& \gamma^2 \tau^{\rho\rho} & 0 & 0 \\
0 &0 & \rho^{-2} \tau^{\phi\phi} &0 \\
0&  0 & 0 & \tau^{-2} \tau^{\eta\eta}
\end{array}
\right),
\end{eqnarray}
\begin{eqnarray}
& & \pi_{\alpha\beta} = g_{\alpha\mu}\, \pi^{\mu\nu} g_{\nu\beta}\nonumber\\[2ex]
& &=\left(
\begin{array}{cccc}
1 &0 &0 &0 \\
0 &-1 &0 &0 \\
0 &0 &-\rho^2 &0 \\
0 &0 &0 &-\tau^2
\end{array}
\right)
\left(
\begin{array}{cccc}
\gamma^2 v_\rho^2\tau^{\rho\rho} & \gamma^2 v_\rho \tau^{\rho\rho}& 0 &0 \\
\gamma^2 v_\rho\tau^{\rho\rho}& \gamma^2 \tau^{\rho\rho} & 0 & 0 \\
0 &0 & \rho^{-2} \tau^{\phi\phi} &0 \\
0&  0 & 0 & \tau^{-2} \tau^{\eta\eta}
\end{array}
\right)
\left(
\begin{array}{cccc}
1 &0 &0 &0 \\
0 &-1 &0 &0 \\
0 &0 &-\rho^2 &0 \\
0 &0 &0 &-\tau^2
\end{array}
\right)\nonumber\\
& &=\left(
\begin{array}{cccc}
\gamma^2 v_\rho^2\tau^{\rho\rho} & \gamma^2 v_\rho \tau^{\rho\rho}& 0 &0 \\
-\gamma^2 v_\rho\tau^{\rho\rho}& -\gamma^2 \tau^{\rho\rho} & 0 & 0 \\
0 &0 & -\tau^{\phi\phi} &0 \\
0&  0 & 0 & -\tau^{\eta\eta}
\end{array}
\right)
\left(
\begin{array}{cccc}
1 &0 &0 &0 \\
0 &-1 &0 &0 \\
0 &0 &-\rho^2 &0 \\
0 &0 &0 &-\tau^2
\end{array}
\right)\nonumber\\
& &=\left(
\begin{array}{cccc}
\gamma^2 v_\rho^2\tau^{\rho\rho} & -\gamma^2 v_\rho \tau^{\rho\rho}& 0 &0 \\
-\gamma^2 v_\rho\tau^{\rho\rho}& \gamma^2 \tau^{\rho\rho} & 0 & 0 \\
0 &0 & \rho^2 \tau^{\phi\phi} &0 \\
0&  0 & 0 & \tau^2 \tau^{\eta\eta}
\end{array}
\right),
\end{eqnarray}
and
\begin{eqnarray}
& & \pi^\mu_\beta = g^{\mu\alpha}\, \pi_{\alpha\beta}\\[2ex]
& &=\left(
\begin{array}{cccc}
1 &0 &0 &0 \\
0 &-1 &0 &0 \\
0 &0 &-1/\rho^2 &0 \\
0 &0 &0 &-1/\tau^2
\end{array}
\right)
\left(
\begin{array}{cccc}
\gamma^2 v_\rho^2\tau^{\rho\rho} & -\gamma^2 v_\rho \tau^{\rho\rho}& 0 &0 \\
-\gamma^2 v_\rho\tau^{\rho\rho}& \gamma^2 \tau^{\rho\rho} & 0 & 0 \\
0 &0 & \rho^2 \tau^{\phi\phi} &0 \\
0&  0 & 0 & \tau^2 \tau^{\eta\eta}
\end{array}
\right)\nonumber\\
& &=\left(
\begin{array}{cccc}
\gamma^2 v_\rho^2\tau^{\rho\rho} & -\gamma^2 v_\rho \tau^{\rho\rho}& 0 &0 \\
\gamma^2 v_\rho\tau^{\rho\rho}& -\gamma^2 \tau^{\rho\rho} & 0 & 0 \\
0 &0 & -\tau^{\phi\phi} &0 \\
0&  0 & 0 & -\tau^{\eta\eta}
\end{array}
\right).
\end{eqnarray}

\subsection{The first term of Eq. (\ref{Trans1})}
The first term of Eq. (\ref{Trans1}) can be expanded as
\begin{eqnarray}
\tau_{\pi} \Delta^{\mu\alpha}\Delta^{\nu\beta} D \pi_{\alpha\beta}
&=& \tau_{\pi}(g^{\mu\alpha}-u^\mu u^\alpha)(g^{\nu\beta}-u^\nu
u^\beta)D\pi_{\alpha\beta}\nonumber\\
&=& \tau_{\pi}g^{\mu\alpha}g^{\nu\beta}D\pi_{\alpha\beta}
-\tau_{\pi}g^{\mu\alpha}u^\nu u^\beta D\pi_{\alpha\beta}\nonumber\\
&-& \tau_{\pi} g^{\nu\beta}u^\mu u^\alpha D\pi_{\alpha\beta}
+\tau_{\pi} u^\mu u^\alpha u^\nu u^\beta D\pi_{\alpha\beta}.
\end{eqnarray}
With the relations $Dg_{\kappa\beta}= Dg^{\kappa\beta} = 0$ and the
orthogonality $u_\mu \pi^{\mu\nu} = u^\mu \pi_{\mu\nu} = 0$, we have
\begin{equation}
g^{\mu\alpha}g^{\nu\beta}D\pi_{\alpha\beta}=g^{\nu\beta}(Dg^{\mu\alpha}
\pi_{\alpha\beta})=g^{\nu\beta}D \pi^\mu_\beta\,,
\end{equation}
\begin{equation}
g^{\mu\alpha}u^\nu u^\beta D \pi_{\alpha\beta} = g^{\mu\alpha}u^\nu
[(Du^\beta\pi_{\alpha\beta})- (D u^\beta)\,\pi_{\alpha\beta}]
= -g^{\mu\alpha} u^\nu (D u^\beta)\, \pi_{\alpha\beta}\,,
\end{equation}
\begin{equation}
g^{\nu\beta}u^\mu u^\alpha D \pi_{\alpha\beta} = g^{\nu\beta}u^\mu
[(Du^\alpha\pi_{\alpha\beta})- (Du^\alpha)\,\pi_{\alpha\beta}]
= -g^{\nu\beta} u^\mu (Du^\alpha)\,\pi_{\alpha\beta}\,,
\end{equation}
\begin{equation}
u^\mu u^\nu u^\alpha u^\beta D\pi_{\alpha\beta} = u^\mu u^\nu
[u^\alpha (Du^\beta\pi_{\alpha\beta})-(Du^\beta)\,
u^\alpha \pi_{\alpha\beta}]=0\,.
\end{equation}
So, the first term of Eq. (\ref{Trans1}) can be expressed as
\begin{equation}
\tau_\pi\Delta^{\mu\alpha}\Delta^{\nu\beta} D {\pi}_{\alpha\beta} =
\tau_\pi(\,g^{\nu\beta} D\pi^\mu_\beta\,+\,I^{\mu\nu}\,)\,,
\end{equation}
where
\begin{equation}
I^{\mu\nu} \equiv g^{\mu\alpha} u^\nu (D u^\beta)\, \pi_{\alpha\beta}
+ g^{\nu\beta} u^\mu (Du^\alpha)\,\pi_{\alpha\beta}\,.
\end{equation}
For the azimuth-symmetric Bjorken cylinder, $u^\phi=0$ and $u^\eta=
0$ at $z=0$, so $I^{\phi\phi}=I^{\eta\eta}=0$.

Therefore, for $\mu=\nu=\rho$ we have
\begin{eqnarray}
\Delta^{\mu\alpha}\Delta^{\nu\beta} D {\pi}_{\alpha\beta}
&=&g^{\rho\beta}D\pi^\rho_\beta\,+\,I^{\rho\rho}
=D\gamma^2 \tau^{\rho\rho}\,+\, I^{\rho\rho}\nonumber\\
&\hspace*{-44mm}=&\hspace*{-23mm} \gamma^3\Big(\frac{\partial}
{\partial\tau}\tau^{\rho\rho} + v_\rho \frac{\partial}{\partial
\rho} \tau^{\rho\rho}\Big)\,+\, 2\gamma (D\gamma)\,\tau^{\rho\rho}\,
+\, I^{\rho\rho},
\end{eqnarray}
where
\begin{eqnarray}
I^{\rho\rho} &=& g^{\rho\alpha} u^\rho (D u^\beta)\, \pi_{\alpha\beta}
+ g^{\rho\beta} u^\rho (Du^\alpha)\,\pi_{\alpha\beta}
= g^{\rho\alpha} u^\rho (D u^\beta)\, \pi_{\alpha\beta}
+ g^{\rho\alpha} u^\rho (Du^\beta)\,\pi_{\beta\alpha}\nonumber\\
&=& 2 g^{\rho\alpha} u^\rho (D u^\beta)\, \pi_{\alpha\beta}
= 2 g^{\rho\rho} u^\rho (Du^\beta) \pi_{\rho\beta}
= -2u^\rho [(D\gamma)\pi_{\rho\tau}+(D\gamma v_\rho)
\pi_{\rho\rho}]\nonumber\\
&=& -2 \gamma v_\rho \big[-(D\gamma) \gamma^2 v_\rho \tau^{\rho\rho}
+ (D\gamma v_\rho) \gamma^2 \tau^{\rho\rho} \big]
= -2 \gamma^4 v_\rho \tau^{\rho\rho} (D v_\rho).
\end{eqnarray}
With the relations,
\begin{equation}
\gamma=\frac{1}{\sqrt{1-v_\rho^2}},~~~~
v_\rho \frac{\partial(\gamma v_\rho)}{\partial \tau}=\frac{\partial
\gamma}{\partial \tau},~~~~
v_\rho \frac{\partial(\gamma v_\rho)}{\partial \rho}=
\frac{\partial \gamma}{\partial \rho},
\end{equation}
we have
\begin{eqnarray}
I^{\rho\rho} &=& -2 \gamma^3 \tau^{\rho\rho} [v_\rho (D\gamma v_\rho)
-v_\rho^2 (D\gamma)]=-2 \gamma^3 \tau^{\rho\rho} \Big[ v_\rho \gamma
\Big(\frac{\partial (\gamma v_\rho)}{\partial \tau} + v_\rho \frac{\partial
(\gamma v_\rho)}{\partial \rho} \Big) -v_\rho^2 (D\gamma)\Big]\nonumber\\
&=& -2 \gamma^3 \tau^{\rho\rho} \Big[ \gamma \Big(\frac{\partial
\gamma}{\partial \tau} + v_\rho \frac{\partial \gamma}{\partial \rho} \Big)
-v_\rho^2 (D\gamma) \Big] = -2 \gamma \tau^{\rho\rho} D \gamma,
\end{eqnarray}
and
\begin{eqnarray}
\Delta^{\mu\alpha}\Delta^{\nu\beta} D {\pi}_{\alpha\beta}
=\gamma^3\Big(\frac{\partial}{\partial\tau}\tau^{\rho\rho} +
v_\rho \frac{\partial}{\partial \rho} \tau^{\rho\rho}\Big)\,,
\end{eqnarray}
for $\mu=\nu=\rho$.

For $\mu=\nu=\phi$ and $\mu=\nu=\eta$, we have
\begin{equation}
\Delta^{\mu\alpha}\Delta^{\nu\beta} D {\pi}_{\alpha\beta} =
\frac{1}{\rho^2} D \tau^{\phi\phi} = \frac{\gamma}{\rho^2}
\Big(\frac{\partial}{\partial\tau}\tau^{\phi\phi} + v_\rho
\frac{\partial}{\partial \rho}\tau^{\phi\phi}\Big)\,,
\end{equation}
and
\begin{equation}
\Delta^{\mu\alpha}\Delta^{\nu\beta} D {\pi}_{\alpha\beta} =
\frac{1}{\tau^2} D \tau^{\phi\phi} = \frac{\gamma}{\tau^2}
\Big(\frac{\partial}{\partial\tau}\tau^{\eta\eta} + v_\rho
\frac{\partial}{\partial \rho}\tau^{\eta\eta}\Big)\,.
\end{equation}

\subsection{The third term of Eq. (\ref{Trans1})}
For the third term of Eq. (\ref{Trans1}), we have
\begin{eqnarray}
\nabla^{<\mu}u^{\nu>}&\equiv &
\nabla^{(\mu}u^{\nu)}-\frac{1}{3}\Delta^{\mu\nu}d_\lambda
u^{\lambda} =\Big[\frac{\nabla^\mu u^\nu +\nabla^\nu
u^\mu}{2}\Big]-\frac{1}{3}\Delta^{\mu\nu}\Theta. \label{T.6}
\end{eqnarray}
With the relations
\begin{equation}
v_{\rho}\frac{\partial(\gamma v_{\rho})}{\partial \tau} =
\frac{\partial \gamma}{\partial \tau},~~~~~~\frac{\partial (\gamma
v_{\rho})}{\partial \rho} = \frac{1}{v_{\rho}}\frac{\partial
\gamma}{\partial \rho}\,,
\end{equation}
we have
\begin{eqnarray}
\nabla^{\rho}u^{\rho}=-\gamma ^2 \Big[\frac{\partial\gamma
}{\partial\tau}+\frac{\partial(\gamma v_{\rho})}{\partial\rho}\Big]
\equiv -\gamma ^2\theta\,,
\end{eqnarray}
and
\begin{eqnarray}
\nabla^{<\rho}u^{\rho>} = \gamma^2 \Big(-\theta + \frac{1}{3}\Theta
\Big) = \gamma^2\sigma^{\rho\rho}\,,
\end{eqnarray}
where
\begin{eqnarray}
\theta=\frac{\partial\gamma}{\partial \tau} +\frac{\partial
(\gamma\,v_\rho)}{\partial \rho}\,.
\end{eqnarray}
Similarly, we have
\begin{equation}
\nabla^\phi u^\phi = \frac{1}{\rho^2} \Big(\frac{-\gamma
v_\rho}{\rho}\Big)\,,
\end{equation}
\begin{equation}
\nabla^{<\phi}u^{\phi>} = \frac{1}{\rho^2} \Big(\frac{-\gamma
v_\rho}{\rho}+\frac{1}{3}\Theta \Big) =
\frac{1}{\rho^2}\sigma^{\phi\phi}\,,
\end{equation}
\begin{equation}
\nabla^\eta u^\eta=\frac{1}{\tau^2}\Big(\frac{-\gamma
}{\tau}\Big)\,,
\end{equation}
\begin{equation}
\nabla^{<\eta}u^{\eta>}= \frac{1}{\tau^2} \Big(\frac{-\gamma}{\tau}
+ \frac{1}{3}\Theta\Big)=\frac{1}{\tau^2}\sigma^{\eta\eta}\,,
\end{equation}
\begin{equation}
\nabla^\tau u^\tau =-\gamma ^2v^2_\rho \Big(\frac{\partial\gamma
}{\partial\tau}+\frac{\partial\gamma v^2_\rho}{\partial\rho} \Big] =
- \gamma^2 v^2_\rho \theta\,,
\end{equation}
\begin{equation}
\nabla^{<\tau}u^{\tau>} = \gamma^2v^2_\rho \Big(-\theta +
\frac{1}{3} \Theta \Big) = \gamma^2v^2_\rho \sigma^{\tau\tau}\,,
\end{equation}
\begin{equation}
\nabla^\tau u^\rho = \nabla^\rho u^\tau = -\gamma^2 v_\rho \theta\,,
\end{equation}
\begin{equation}
\nabla^{<\tau}u^{\rho>} = \nabla^{<\rho} u^{\tau>} = \gamma^2 v_\rho
\Big(-\theta + \frac{1}{3}\Theta\Big)=\gamma^2 v_\rho \sigma^{\tau
\rho}\,.
\end{equation}

\subsection{The last term of Eq. (\ref{Trans1})}
The last term of Eq(\ref{Trans1}) can be written as
\begin{eqnarray}
\widetilde{\eta}\,T d_{\lambda} \bigg(
\frac{\tau_{\pi}u^{\lambda}}{\widetilde{\eta}} T \bigg) =
\widetilde{\eta}\,T \bigg[\frac{\tau_\pi}{\widetilde{\eta}\,
T}d_\lambda u^\lambda + u^\lambda d_\lambda \Big(\frac{\tau_\pi
}{\widetilde{\eta}\,T}\Big) \bigg]=\tau_\pi \Theta +
\widetilde{\eta}\,T D \Big(\frac{\tau_\pi}{\widetilde{\eta}\,
T}\Big)\,.
\end{eqnarray}\\

Now, one can obtain Eqs. (\ref{T.rr}) -- (\ref{T.eta}) from the
above derivations.


\end{document}